\documentclass[aps,prl,twocolumn,eqsecnum,amssymb,amsmath,showpacs,a4paper, superscriptaddress]{revtex4-2}

\usepackage{graphicx}
\usepackage{amsfonts}
\usepackage{amsmath}

\usepackage[dvipsnames]{xcolor}
\usepackage[unicode]{hyperref}
\hypersetup{colorlinks=true, citecolor=MidnightBlue,
            linkcolor=MidnightBlue, urlcolor=MidnightBlue, linktocpage=true}
            
\usepackage{units}
\usepackage{color}
\usepackage{dcolumn}
\usepackage{bm}
\usepackage{float}
\usepackage{cleveref}
\usepackage[normalem]{ulem}
\usepackage{appendix}
\usepackage{mathtools}
\usepackage{esvect}

\usepackage[colorinlistoftodos, size=tiny, bordercolor=white]{todonotes}
\usepackage{comment}

\graphicspath{ {images/} }
\usepackage{mathrsfs}
\usepackage{amssymb}
\usepackage{dsfont}
\usepackage{enumitem}
\usepackage{gensymb}
\usepackage{bm}

\usepackage{mathtools}
\usepackage{chngcntr}
\counterwithout{equation}{section}

\makeatletter
\let\cat@comma@active\@empty
\makeatother

\usepackage{xr}
\usepackage{orcidlink}
\newcommand{\orcid}[1]{\href{https://orcid.org/#1}{\includegraphics[width=10pt]{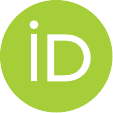}}}

\begin{document}

\title{Spectroscopy of analogue black holes using simulation-based inference}

\author{Leonardo Solidoro \orcid{0009-0001-9147-445X}}
\email{leonardo.solidoro@nottingham.ac.uk}
\affiliation{School of Mathematical Sciences, University of Nottingham, University Park, Nottingham, NG7 2RD, UK}
\affiliation{Nottingham Centre of Gravity, University of Nottingham,
University Park, Nottingham NG7 2RD, UK}

\author{Sebastian H.\,V\"olkel \orcid{0000-0002-9432-7690}
}
\email{sebastian.voelkel@uni-tuebingen.de}
\affiliation{Theoretical Astrophysics, IAAT, University of T\"ubingen, Auf der Morgenstelle 10, D-72076 T\"ubingen, Germany}
\affiliation{Max Planck Institute for Gravitational Physics (Albert Einstein Institute), Am M\"uhlenberg 1, 
D-14476 Potsdam, Germany}

\author{Silke Weinfurtner \orcid{0000-0001-7642-5476}}
\email{silke.weinfurtner@manchester.ac.uk}
\affiliation{School of Mathematical Sciences, University of Nottingham, University Park, Nottingham, NG7 2RD, UK}
\affiliation{Nottingham Centre of Gravity, University of Nottingham,
University Park, Nottingham NG7 2RD, UK}
\affiliation{Centre for the Mathematics and Theoretical Physics of Quantum Non-Equilibrium Systems, University of Nottingham, Nottingham, NG7 2RD, UK}
\affiliation{Department of Physics and Astronomy, University of Manchester, Manchester M13 9PL, UK}
\affiliation{Photon Science Institute, Alan Turing Building, University of Manchester, Manchester M13 9PY, UK}

\date{\today}
\begin{abstract}
The emergence of precision gravity simulators in quantum and fluid systems is opening new avenues for probing curved-spacetime physics and black-hole phenomenology under controlled laboratory conditions. 
In parallel, advances in understanding how fundamental physics can be probed in the spectral signatures of black holes and exotic compact objects motivate the development of modern spectroscopic techniques within analogue-gravity experiments. 
In this work, we model the spectral properties of analogue black holes sourced by broadband stochastic noise, a crucial aspect in realistic experiments that poses substantial challenges for established data-analysis techniques. 
Using simulation-based inference, we demonstrate that the physical parameters encoded in noisy spectra can be reliably extracted, showing that these techniques provide a powerful tool for studying both spacetime properties and boundary effects in gravity simulators. 
\end{abstract}
\maketitle

\section{Introduction}\label{sec:intro}

Black holes represent some of the most extreme environments in the universe, challenging our current understanding of fundamental physics. While modern experiments can probe the strong gravity effects of black holes by measuring relativistic orbits~\cite{GRAVITY:2018ofz,GRAVITY:2020gka}, shadows~\cite{EventHorizonTelescope:2019dse,EventHorizonTelescope:2022wkp}, and gravitational waves~\cite{LIGOScientific:2016aoc,LIGOScientific:2025obp}, many theoretical predictions remain inaccessible for astrophysical black holes. 
The black hole spectroscopy programme, in particular, aims to probe fundamental physics by detecting the characteristic quasinormal modes (QNMs) from perturbed compact objects. See Refs.~\cite{Regge:1957td,Zerilli:1970se,Vishveshwara:1970zz,Teukolsky:1973ha,Chandrasekhar:1975zza} for some of the pioneering works and Refs.~\cite{Kokkotas:1999bd,Nollert:1999ji,Berti:2009kk,Konoplya:2011qq,berti2025spectroscopy} for topical reviews.

Phenomena previously considered inaccessible in curved spacetime and black hole physics have been successfully observed in gravity simulators. 
These are systems whose perturbations behave as propagating on a curved spacetime~\cite{unruh1981,barcelo2011,Schutzhold2002}. 
For more than one decade, controlled tabletop experiments of such systems have permitted the measurement of black-hole phenomena such as Hawking radiation~\cite{weinfurtner2011,steinhauer2014,steinhauer2016,euve2016,munoz2019observation,steinhauer2022conf} and rotational superradiance~\cite{visser2005,Torres2017,solnyshkov2019,cromb2020,braidotti2022}. 
Notably, the prospects of studying QNM  emission in the laboratory have been at the centre of analogue black holes studies \cite{Berti2004,Cardoso2004,Dolan2009,torres2018,Dolan2012}. 
predicted in polariton superfluids \cite{jacquet2023,Guerrero:2025kdn}, optical solitons~\cite{Burgess:2023pny,Kranas:2025taq}, in hydrodynamic systems~\cite{patrick2018,torres2020}, superfluids~\cite{svancara2024,Smaniotto:2025hqm}, cold-atomic-gases~\cite{kolobov2021observation,geelmuyden2022}. 
These findings establish gravity simulators as powerful platforms for black hole spectroscopy in non-standard settings, where dispersion~\cite{patrick2020, patrick2025}, confinement~\cite{Solidoro:2024yxi}, and angular-momentum quantisation~\cite{geelmuyden2022,patrick2022} enrich the phenomenology, enabling controlled access to regimes beyond the reach of observations and current numerical approaches.

Besides the underlying curved spacetime geometry, the dynamics of perturbations in a gravity simulator depend on the specific boundary conditions, which can depart from the purely radiative ones of gravitational black holes. Similar deviations also arise in models of horizonless ultra compact objects or new physics on the horizon scale~\cite{Schwarzschild:1916ae,Bronnikov:1973fh,Ellis:1973yv,Mazur:2001fv,Mazur:2004fk,Damour:2007ap,Almheiri:2012rt,Cardoso:2016oxy,Bueno:2017hyj,Cardoso:2019apo,Coates:2019bun,Maggio:2020jml}; see Ref.~\cite{Cardoso:2019rvt} for a review. 
In such systems, partially-reflective boundary conditions near the would-be horizon give rise to qualitatively different QNMs (including trapped modes)~\cite{3f29f6b9-052e-309b-94f7-d9b7c3e0aaca,Kokkotas:1994an}, and novel features in the time-domain (echoes)~\cite{Kokkotas:1995av,Ferrari:2000sr,Tominaga:1999iy,Cardoso:2016rao,Maselli:2017tfq,Testa:2018bzd}. 
Furthermore, the impact of perturbations in the scattering potential in the exterior of the black hole has recently received attention in spectral stability studies~\cite{Jaramillo:2020tuu,Daghigh:2020jyk,Cheung:2021bol,Destounis:2021lum,Berti:2022xfj,Rosato:2024arw,Boyanov:2022ark}, which extend pioneering works on the significance of QNM excitations~\cite{Nollert:1996rf,Leung:1997was,Nollert:1998ys}. 

Another feature in analogue models of compact objects, which is fundamental to experiments, is the presence of broadband noise driving the system's dynamics. 
It allows for the direct observation of the system's QNMs in the frequency domain~\cite{Solidoro:2024yxi,Smaniotto:2025hqm}. 
However, the stochastic nature of the noise driving makes the modelling inherently non-deterministic, which distinguishes the spectroscopy of analogues from its gravitational counterpart, where the ringdown signal of a binary merger is a clean deterministic process, and noise originates from the detector. 
This significantly challenges the application of well-established Bayesian data-analysis approaches used in gravitational-wave applications, where detector noise properties can be characterized in the absence of a signal. 

In this work, we overcome this challenge using simulation-based inference (SBI), a machine-learning-based probabilistic framework tailored to settings where the data are intrinsically noisy and a likelihood formulation is intractable. SBI operates directly on forward simulations and does not require an explicit separation between signal and noise~\cite{doi:10.1073/pnas.1912789117,deistler2025simulationbasedinferencepracticalguide}, making it ideally suited to analogue systems driven by stochastic forcing. Previous applications to gravitational-wave inference~\cite{Dax:2021tsq,Dax:2022pxd,Dax:2024mcn,Crisostomi:2023tle} and neutron-star universal relations~\cite{Kruger:2026bnz} have showcased the potential of SBI.

We focus on two theoretical models used to characterize analogue black holes in experiments. 
First, we consider a P\"oschl-Teller potential barrier within a finite domain and mechanical noise, as first introduced in Ref.~\cite{Solidoro:2024yxi}. We then turn to noise-driven shallow-water waves on a vortex flow, which provides a minimal model for hydrodynamical black-hole simulators such as~\cite{torres2020,Smaniotto:2025hqm}.

Using numerical solutions of stochastic differential equations as training data for SBI, we demonstrate that neural posterior estimation (NPE) enables the successful extraction of key parameters. 
We also report the reconstruction of the system's Green's function in the P\"oschl-Teller case, and conjecture its reconstruction in the shallow-water case, which allows a full characterisation of the systems and boundary conditions from a single dataset. This positions SBI as a powerful tool for spectral analysis in experiments with limited statistics, intrinsically uncertain initial conditions, or limited control over state preparation.

\section{Methods}\label{sec:methods}

\emph{Analogue black hole model}---
We consider the one-dimensional wave equation with a potential $V(x)$ and the time-dependent sourcing term $I(x,t)$
\begin{align}
    \label{eq:WaveEq}
    -\frac{1}{c^2}\partial_t^2 \psi+ \partial_x^2 \psi - V(x)\psi = I(t,x)\,,
\end{align}
where $c$ denotes the wave speed, which we set to unity. 
Equations of the form Eq.~\eqref{eq:WaveEq} are central both to black-hole perturbation theory and to analogue gravity systems. 
In the black-hole case, they arise after separating variables in the perturbation equations, with $x$ denoting the so-called tortoise coordinate. 
In analogue setups, similar structures arise from linearized models of waves on a background flow or condensate~\cite{barcelo2011, torres2018,patrick2025}. 

Taking the Laplace transform of Eq.~\eqref{eq:WaveEq} with respect to time yields a sourced Helmholtz equation,
\begin{align}
    \label{scrEq}
\partial_x^2\phi + \big[\omega^2 - V(x)\big]\phi = J(\omega,x),
\end{align}
for $\phi(\omega,x) = \mathcal{L}[\psi(t,x)]$, with $J(\omega,x)$ including both the Laplace transform of the source term and the initial condition contributions \cite{Kuntz:2025gdq}.
The equation can be solved using the variation of constants method, which requires introducing the Green's function $G(x,x_s,\omega)$ satisfying Eq.~\eqref{scrEq} with $J = \delta(x-x_s)$ and appropriate boundary conditions. 
Then, the solution for any given source term is given by
\begin{align}
    \label{resGreen}
    \phi(\omega,x) = \int d x_s G(x,x_s,\omega)  J(\omega,x_s)\,.
\end{align}
The full solution of Eq.~\eqref{eq:WaveEq} is therefore reduced to calculating the inverse Laplace transform of $G(x,x_s,\omega)$. 
Although the problem can be solved numerically, understanding the analytical properties of $G(x,x_s,\omega)$ is central for black hole spectroscopy. In particular, for asymptotically vanishing potentials, $V(x)\to 0$ as $x\to \pm \infty$, its poles in the complex frequency plane determine the discrete set of decaying oscillations, known as QNMs \cite{Leaver:1985ax,Leaver:1986gd}. 
These correspond to solutions satisfying radiative boundary conditions
\begin{align}
\label{BCsOpen}
\lim_{x\to \pm \infty}\phi(x) \,\propto\, e^{\pm i \omega x},
\end{align}
i.e., purely outgoing waves on both sides of the potential.

In experiments, radiative boundary conditions Eq.~\eqref{BCsOpen} are rarely exact, as waves interact with finite boundaries and interfaces that induce partial reflection. 
To capture these effects, we impose imperfect radiative boundary conditions at an inner point $x_l$ and an outer point $x_r$. 
At each boundary, incident waves are reflected with amplitudes multiplied by $\varepsilon_l$ and $\varepsilon_r$, giving the asymptotic behaviour
\begin{align}
\label{BCs}
    \phi\propto \begin{cases}
    e^{-i\omega x} + \varepsilon _le^{+i\omega (x-2x_l)} \,,&\,\,x\to x_l \,,\\
    \varepsilon_re^{-i\omega (x-2x_r)} + e^{+i\omega x}\,,&\,\,x\to x_r\,.
    \end{cases} 
\end{align}
The parameters $\varepsilon_{l,r}$ quantify the reflectivity at the left and right boundaries, with the amplitude of the incident wave at each boundary normalized to unity.

A generic Robin boundary condition $\alpha \phi + \beta \phi' = \gamma$ can be recovered by choosing
\begin{align}
\label{robin_bound}
    \varepsilon_{l/r} = \left(\frac{i\omega \beta -\alpha}{i \omega \beta + \alpha }\right) - \gamma\,.
\end{align}
In particular, for Neumann ($\alpha = \gamma = 0$) and Dirichlet  ($\beta = \gamma = 0$) boundary conditions one gets $\varepsilon_{l/r} = 1,-1$ respectively. One recovers the radiative boundary condition with $\varepsilon_{l/r}=0$. 

Since the Green’s function $G(x,x_s,\omega)$ encodes the full linear response, any departure from ideal radiative behaviour appears directly in its pole structure. 
Partial reflection at boundaries modifies the frequencies and widths of these poles, allowing the effective boundary condition, in principle, to be inferred experimentally \cite{Solidoro:2024yxi}.

\emph{Noise spectroscopy}--- In realistic experiments, systems are naturally driven by broadband stochastic (e.g., mechanical or thermal) noise, which constitutes a natural probe of the Green’s function. 
The spectral features of the system are captured by the power spectral density (PSD) of the field 
\begin{align}
    S_\phi(\omega,x) = \langle |\phi(\omega,x)|^2\rangle\,,
\end{align}
i.e., the squared Fourier transform of the field averaged over many noise realisations. For a linear system Eq.~\eqref{eq:WaveEq}, the PSD of the field Eq.~\eqref{resGreen}  factorises as \cite{newland2005introduction}
\begin{equation}
    S_\phi = |G(x,x_s,\omega)|^2 S_J\,, 
\end{equation}
where the PSD of the source $S_J =  \langle |J|^2\rangle$ is constant in frequency under white-noise driving. 
The observed PSD therefore gives immediate access to the pole structure of the Green’s function and, through it, to the underlying boundary conditions. 

Nonetheless, in many experimental settings, the available statistics are inherently limited: acquisition times may be short, noise realisations cannot always be repeated, and slow drifts in the background conditions prevent long ensemble averages. 
When only a single or a few stochastic realisations are accessible, the PSD becomes a noisy estimator of $|G|^2$, and the resonant structure is difficult to extract by direct fitting. 
Under these circumstances, Bayesian methods provide a principled way to infer parameters directly from the data while accounting for uncertainty.

\emph{Simulation-based inference---} 
The stochastic differential equation used to model the data $d$ (spectrum) of our experiments makes the standard application of Bayesian methods, e.g., Markov-chain-Monte-Carlo (MCMC) techniques, challenging. 
In a real experiment, it is, in general, not trivial to construct the likelihood function $p(d|\theta)$. 
However, it is central for Bayesian analysis for obtaining the posterior $p(\theta|d)$ via Bayes' theorem
\begin{align}
p(\theta|d) = \frac{p(d|\theta) p(\theta)}{p(d)}\,,
\end{align}
where $p(\theta)$ is the prior and $p(d)$ the evidence. 
To define the likelihood, one would usually provide the model as a function of the parameters $m(\theta)$ (without noise), and quantify the noise $n$, so that the actual data is described by $d = m(\theta) + n$. 
In many applications, e.g., gravitational wave data analysis, one often encounters approximate distributions, such as stationary Gaussian noise, which allows one to compute the posterior by sampling it using MCMC techniques. 
However, in our case, noise is a central part of the system itself and cannot be measured independently. 

SBI, also known as likelihood-free inference, circumvents the problem of not knowing the likelihood function. 
Its key idea is to use machine-learning techniques (such as normalizing flows) that rely solely on simulations that already include noise, thereby avoiding the explicit separation of model and noise. 
Because the outcome of a stochastic equation for given $\theta$ already includes noise realizations, it is an ideal use case for SBI.
We use the popular \texttt{Python} package \texttt{sbi}~\cite{tejero-cantero2020sbi,tejero_cantero_2022_8192532,BoeltsDeistler_sbi_2025}, which covers several machine-learning techniques. 
For our application, we utilize NPE~\cite{Papamakarios:2016ctj,lueckmann2017flexiblestatisticalinferencemechanistic,greenberg2019automaticposteriortransformationlikelihoodfree,deistler2022truncatedproposalsscalablehasslefree}. 
Providing the noisy spectra for many parameters and noise realizations as training data, NPE allows, after successful training, to sample from an approximate posterior distribution of a new observation.  
Compared to traditional Bayesian methods, NPE is fully amortized. 
The only significant computational cost lies in providing simulations for training, while posterior sampling only takes a few seconds. 
This makes NPE ideally suited for laboratory experiments with comprehensive data acquisition. 
Further details are reported in the Appendix.

\section{Application and Results}\label{sec:app}
\emph{P\"oschl-Teller model}--- \label{sec:app_1}
As our first application, we consider the P\"oschl-Teller potential
\begin{equation}
    V_\mathrm{PT}(x) = V_0\cosh^{-2}[\alpha (x-x_0)]\,,
\end{equation}
where $V_0 >0$ describes the height of the barrier at the maximum, $x_0$ its location, and $\alpha$ scales its width. 
It is often used as a model for qualitative studies of QNMs, e.g., Refs.~\cite{Ferrari:1984ozr,Beyer:1998nu,Price:2017cjr,Cheung:2021bol,Volkel:2022ewm,Nee:2023osy,Kuntz:2025gdq}. Due to its simplicity, it is possible to compute its Green's function analytically. 
This has recently been done in Ref.~\cite{Kuntz:2025gdq} for purely radiative boundary conditions Eq.~\eqref{BCsOpen}, which we generalize to Eq.~\eqref{BCs} in the Appendix. 

In addition to the potential's parameters, the system is characterized by the positions of the left and right boundaries, $x_l$ and $x_r$, the descriptions of the boundary conditions through $\varepsilon_l$ and $\varepsilon_r$, and the injected noise amplitude, $\sigma$. 
Since the system size can typically be measured independently in experiments, we treat $x_r$ as known and fix it to 4. 
Thus, in total, we consider the set of free parameters
\begin{align}
\theta_\mathrm{PT} = [V_0, \alpha, x_0, x_l, \varepsilon_l, \varepsilon_r, \sigma]\,. 
\end{align}

To train the NPE, we generate $10^5$ stochastic wave evolutions and compute the corresponding power spectra. Specifically, we sample the parameters from uniform distributions within given ranges, then we solve the forward model Eq.~\eqref{eq:WaveEq} using finite-difference discretisation in space and 4th-order Runge-Kutta time integration. We take $\psi(x,t=0) = 0$ as the initial condition but generate random noise from the left boundary at each time step for the duration of the simulation. We then Fourier transform the signal after the noise has propagated across the system and take the average near the right boundary. 
Further details are provided in the Appendix. 

\begin{figure}
    \centering
    \includegraphics[width=1\linewidth]{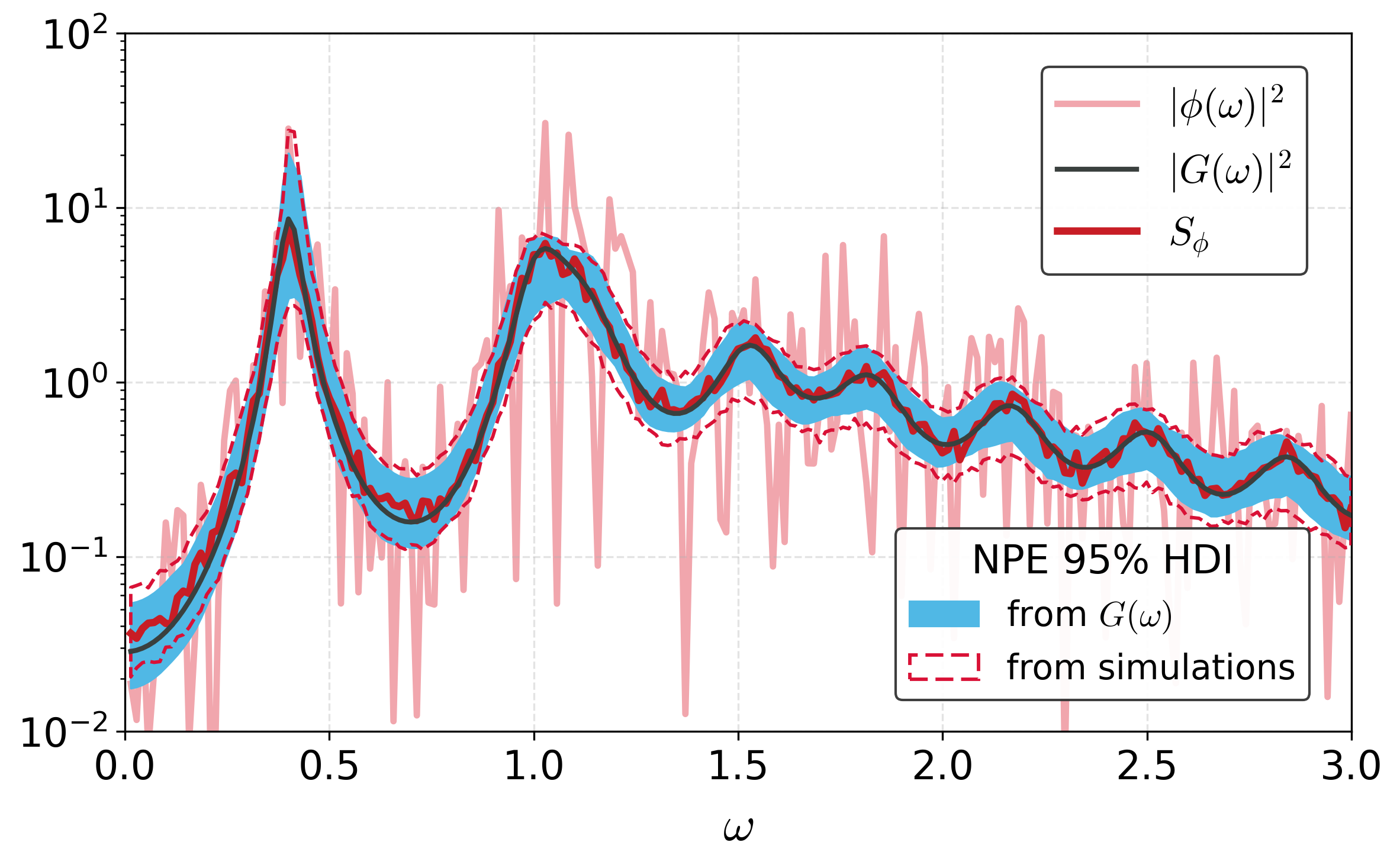}
    \caption{Application of NPE to the spectrum of P\"oschl-Teller noise with $\theta_\mathrm{PT} = [1.2, 0.9, -0.5, -5, 0.8, 0.2, 3]$ and $x_r = 4$.
    The normalized power spectrum obtained from a single noisy wave realisation (light red curve) is shown in log scale and compared with the noise-averaged power spectral density $S_\phi$ (red curve) and the squared modulus of the analytic Green’s function $|G(\omega)|^2$ (dark curve).
    The regions within the blue band (dashed lines) indicate the 95\% credible interval built from the Green's functions (the numerical PSD) with parameters drawn from the posterior distribution estimated by the NPE.}
    \label{fig:spectrum}
\end{figure}

We apply the NPE to a noisy spectrum in Fig.~\ref{fig:spectrum}, where the inference is performed for a signal generated with P\"oschl-Teller parameters $\theta_\mathrm{PT}$ reported in the caption. 
The single noisy realisation (light red curve) produces a spectrum in which the resonant peaks are barely discernible. 
The PSD is obtained by averaging over 50 independent noise realisations with identical $\theta_\mathrm{PT}$ (red curve), which sharpens the peaks, and is in excellent agreement with the analytic Green’s function (black curve). The discrepancy at low frequency is an effect of the limited time evolution. 
This provides important confirmation of our computation, demonstrating the close relation between the spectrum and the system's Green's function. 

Sampling the posterior distribution estimated by the NPE (the full 7-dimensional posterior is shown in Fig.~\ref{fig:PT_corner}, with axis spanning the full prior ranges), one can construct the spectrum's 95\% confidence interval at each frequency by either computing the Green's function (blue area) or by numerically estimating the PSD (red-dashed contour) averaging over 50 realisations for each posterior draw. Crucially, these confidence intervals are obtained from inference on a single noisy spectrum, and therefore provide an estimate of the underlying spectral structure without the need for ensemble averaging over repeated experiments. 
Indeed, we find that the injected parameter values are well represented by the posterior distribution, confirming that the SBI approach enables the reconstruction of the scattering potential and the physical properties of the system from a single noise realisation.

\emph{Shallow-water waves}\label{sec:app_2}--- 
As an application to gravity simulators, we also study the propagation of gravity waves on the free surface of a fluid flow~\cite{Schutzhold2002,torres2018,torres2020,torres2022imp,patrick2018,patrick2020}. We consider an irrotational, inviscid background flow $\vec{v}_0 = \nabla\Phi$, with velocity potential $\Phi$. 
Small perturbations of the potential, $\phi$, describe surface waves. In the shallow-water limit (wavelengths much larger than the fluid depth $h$), $\phi$ satisfies the effective wave equation
\begin{align}
    \label{eq:SW}
    (\partial_t+\mathbf{v}_0\cdot \nabla)(\partial_t+\mathbf{v}_0\cdot \nabla)\phi - c_s^2\nabla^2 \phi = 0\,,
\end{align}
where the propagation speed is set by the fluid depth, $c_s = \sqrt{gh}$. 
A background of particular interest is the draining bathtub vortex, given in polar coordinates $(r,\theta)$ by $\mathbf{v}_0 = -({D}/{r})\hat{r} + ({C}/{r})\hat{\theta}\,,$ where $D$ and $C$ determine the draining and circulating components of the flow, respectively. 
The analogy with black holes arises upon noticing that Eq.~\eqref{eq:SW} is equivalent to the Klein–Gordon equation for a scalar field propagating on an effective curved spacetime with line element
\begin{align}
    \label{metric}
    ds^2 = -c_s^2dt^2 + \left(dr + \frac{D}{r}dt\right)^2 + \left(r d\theta-\frac{C}{r}dt\right)^2\,.
\end{align}
The effective spacetime possesses an acoustic horizon at $r_h = D/c_s$ and an ergosurface at $r_e = \sqrt{C^2 + D^2}/c_s$. 
These structures provide the analogue of a rotating black hole spacetime \cite{visser2005}.
From now on, lengths and times will be scaled with respect to $r_h$ and $r_h/c_s$, respectively (this amounts to setting $c_s = D = 1$). 

In a stationary, axisymmetric system, the field can be de-composed in its Fourier and azimuthal components $\phi = \psi \exp[i(m\theta - \omega t)]$. 
After a standard radial field redefinition (see e.g. \cite{torres2022imp} for details) and the introduction of a tortoise coordinate $x$ defined by $dx = dr/(1-r^{-2})$, the dynamics of each $(\omega,m)$ mode outside the horizon reduces to a one-dimensional wave equation
\begin{align}
    \label{SWeq}
    \partial_x^2 \psi +\left(\widetilde\omega^2 -  V(r)\right)\psi = 0\,,
\end{align}
with $\widetilde\omega = \omega -mC/r^2$ the frequency in the co-moving frame, and the scattering potential given by
\begin{align}
    V(r) =\left( 1 - \frac{1}{r^2}\right)\left(\frac{m^2 - 1/4}{r^2} + \frac{5}{4 r^2}\right)\,.
\end{align}

\begin{figure}
    \centering
    \includegraphics[width=1\linewidth]{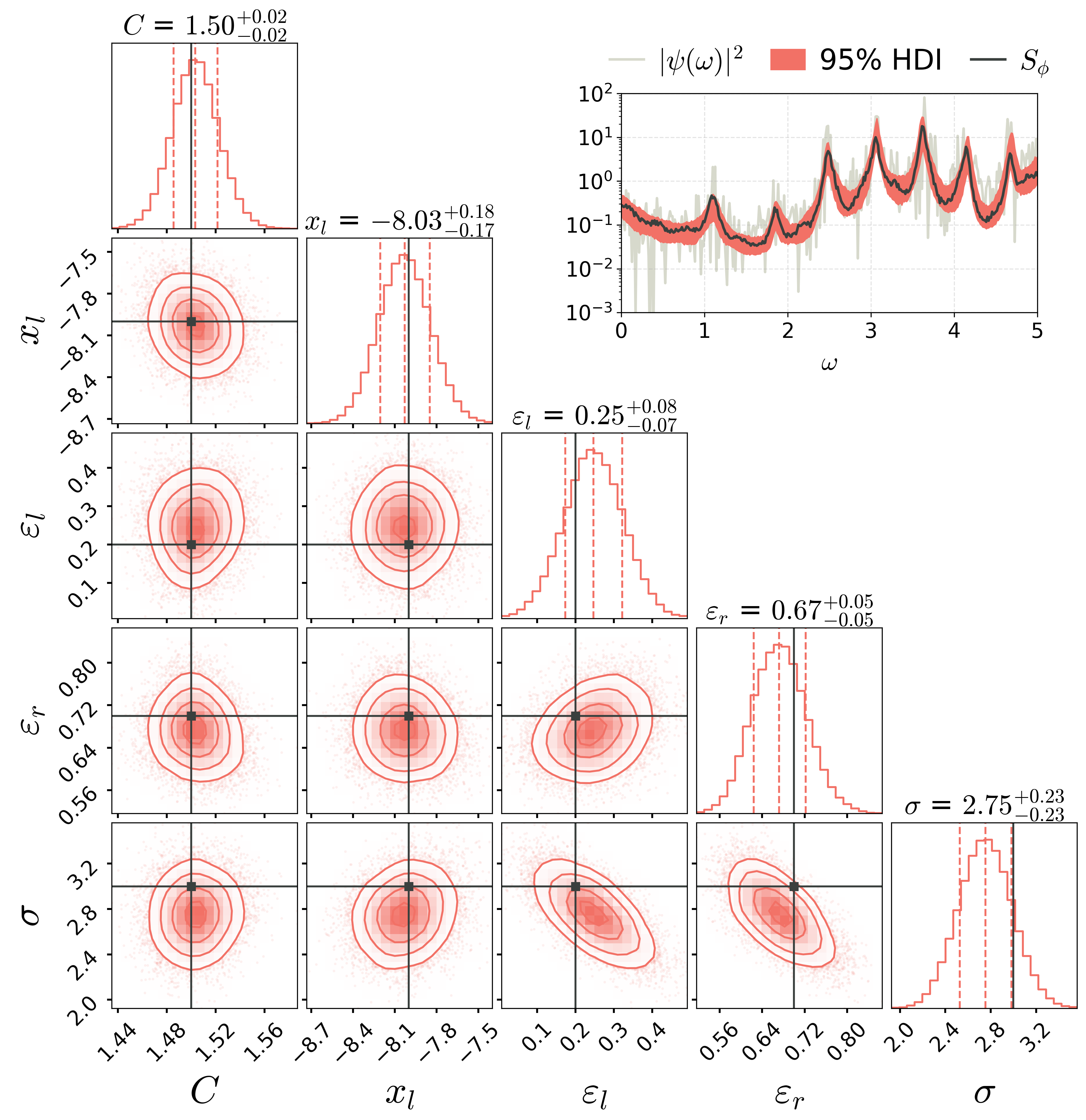}
\caption{Posterior parameter inference for a shallow-water analogue black hole using NPE.
Shown is the posterior distribution of conditioned on a single simulated spectrum with injected parameters $\theta_\mathrm{SW}=[1.5,-8,0.2,0.7,3]$ (black lines).
Diagonal panels display the one-dimensional marginal posteriors with median and 68\% credible intervals, while off-diagonal panels show the corresponding joint distributions.
Inset: the normalized noisy power spectrum (light curve) is compared with the noise-averaged PSD $S_\phi$ (black curve). 
The red band represents the 95\% confidence interval for the spectrum built by averaging over 50 noise realisations for each posterior draw.}
    \label{fig:SW_Corner}
\end{figure}

\begin{figure*}
    \centering
    \includegraphics[width=1.0\linewidth]{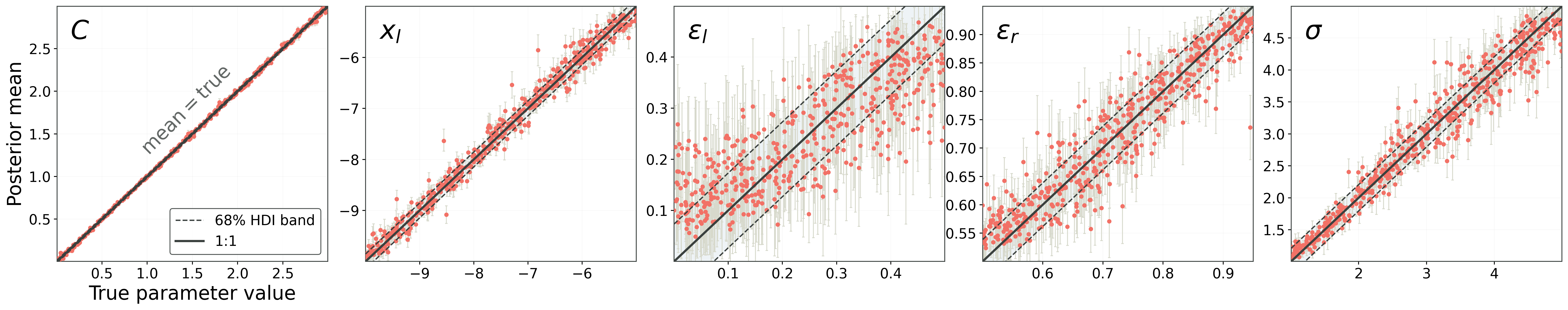}
    \caption
    {Validation of the NPE for shallow-water perturbations with $m=4$ using 500 samples from the test set. 
    The mean of the posterior is plotted against its corresponding true value (orange markers), with 68\% highest-density interval (HDI) error bars shown for every sample. 
    The solid diagonal indicates perfect recovery, while the dashed lines denote the average 68\% HDI width around the identity.
}
    \label{fig:SW_Valid}
\end{figure*}

In this case, scattering of waves with fixed azimuthal number $m$ depends only on the circulation parameter $C$, which in a fluid-dynamical experiment can be independently constrained by non-spectral measurements. 
The free parameters are therefore
\begin{equation}
\theta_\mathrm{SW} = [C, x_l, \varepsilon_l, \varepsilon_r,\sigma] \,.
\end{equation}
We fix $m=4$ and the right boundary $x_r = 6$ and train the NPE with uniform priors $C\in[0,3]$, $x_l\in[-10,-5]$, $\varepsilon_l\in[0,0.5]$ and  $\varepsilon_r\in[0.5,.95]$. The noise variance is sampled from a uniform distribution $\sigma\in [1,5]$.

We apply NPE to obtain the posterior distribution for a single simulated spectrum shown in Fig.~\ref{fig:SW_Corner}, with injected shallow-water parameters $\theta_\mathrm{SW}$ reported in the caption. The posterior has a simple structure (no multimodality) and recovers the injected values within the 68\% credible regions for all parameters.

Notably, the circulation parameter $C$ is tightly constrained, with relative uncertainty of around 1.33\%. This demonstrates the effectiveness of inference from single noise realisations even when compared to strategies based on averaging over repeated experimental runs. While $C$ can, in principle, be externally controlled, reaching comparable constraints on the spectrum through averaging would require percent-level stability across realisations. Residual fluctuations in 
$C$ then propagate through the inference and bias parameters such as $x_l$, rendering the averaging procedure non-trivial. 
In the inset, we show the corresponding 95\% credible interval for the PSD, obtained by averaging over 50 noise realisations for each posterior draw (orange area), together with the noisy spectrum (light curve) and the PSD of the injected parameters when averaging over many noise realizations (black curve). 

To demonstrate the NPE's performance beyond a single example, we also validate the inference on an ensemble of independent simulated spectra. 
Fig.~\ref{fig:SW_Valid} summarises this validation using a representative subset of 500 spectra drawn from the test set. 
For each case, the posterior mean is compared to the true parameter value used in the simulation (orange markers), with error bars indicating the 68\% high-density interval. 
It is evident that $C$ and $x_l$ are tightly constrained, while $\varepsilon_l$ and $\varepsilon_r$ are less. This demonstrates that the spectra are more sensitive to the first two parameters and less to the others. In realistic experiments, multiple 
$m$-modes are simultaneously accessible, and their combined analysis is expected to improve the constraints on the effective boundary conditions significantly.

\section{Conclusions}
\label{sec:conc} 

Experimental advances in gravity simulators of quantum and fluid systems provide unprecedented avenues for probing curved-spacetime physics under controlled laboratory conditions. 
However, recent studies revealed that some of these systems are driven by noise, and can be modelled by stochastic differential equations~\cite{Solidoro:2024yxi,Smaniotto:2025hqm}. 
This poses significant challenges for traditional data analysis methods, which rely on clean models and explicit knowledge of the likelihood function. 

In this work, we demonstrated how SBI techniques~\cite{doi:10.1073/pnas.1912789117,deistler2025simulationbasedinferencepracticalguide}, in particular NPE~\cite{Papamakarios:2016ctj,lueckmann2017flexiblestatisticalinferencemechanistic,greenberg2019automaticposteriortransformationlikelihoodfree,deistler2022truncatedproposalsscalablehasslefree}, overcome the likelihood problem, and can be successfully used for spectroscopic studies of analogue black holes. 
We focused on two theoretical models to describe the physical process of wave scattering due to a stochastic noise source. 
We first studied the P\"oschl-Teller potential in a finite cavity, as introduced in Ref.~\cite{Solidoro:2024yxi}. 
After training of NPE, we injected a noisy spectrum and successfully recovered the associated 7-dimensional posterior distribution. 
Moreover, we demonstrated that applying NPE enables direct inference of the system's Green's function. 
Then, we repeated the analysis for the shallow-water waves model, because it is closer to actual experiments and directly contains the physical parameters of the analogue system. 
Calibration analyses reported in the Appendix suggest that NPE reliably estimates posterior distributions for both systems. 

In real fluid experiments, determining surface-wave boundary conditions is intrinsically challenging. The Robin parameters $\alpha, \beta, \gamma$
in Eq.~\eqref{robin_bound} encode the combined effects of viscosity, surface tension, geometry, and container material~\cite{miles1967,cocciaro1993,jiang2004,kidambi2009,viola2018,bongarzone2021}, and a predictive first-principles description is generally unavailable, and the effective boundary parameters must be inferred empirically for each specific fluid–boundary configuration. Near the vortex core (here the left boundary), large flow velocities, strong free-surface curvature, and concentrated vorticity further complicate the wave dynamics. From the perspective of analogue gravity, these effects are expected to prevent the formation of an ideal one-way membrane, replacing the effective horizon with a partially transmitting inner boundary whose reflectivity cannot be computed from first principles. 
In this respect, SBI provides a viable framework for constraining effective reflectivities, enabling a quantitative characterization of boundary effects directly from spectral data. 

Complementary developments concern the reconstruction of the effective potential of analogue black holes from transmission or reflection coefficients~\cite{Albuquerque:2023lzw,Albuquerque:2024cwl}. 
Based on extending semi-analytic methods~\cite{lieb2015studies,MR985100,1980AmJPh..48..432L,2006AmJPh..74..638G,Volkel:2017kfj,Volkel:2018hwb,Volkel:2019ahb,Albuquerque:2024xol}, the approach does not rely on parametrized models, but does, not yet, account for measurement uncertainties. 
Whether it is possible to combine SBI with the semi-analytic method would be an interesting question for future work. 
Inferring analogue black hole metrics with traditional MCMC methods from the time evolution of perturbations~\cite{Albuquerque:2025eny} might apply to optical fibres and solitons~\cite{Burgess:2023pny,Kranas:2025taq}. 

Possible future extensions of our work are the application of the SBI framework to real data from analogue gravity experiments. 
Real data is, in general, expected to contain imprints of imperfections of the experimental setup and possibly more involved physical effects, e.g., dissipation or deviations from the shallow-water approximation. 
Therefore, it will be an interesting and challenging task to verify to what extent the here considered models may need to be extended to yield robust results for an SBI application, thereby avoiding model misspecification. 
Since some, but not all, of the parameters can also be measured independently, one could directly test the SBI predictions.

\acknowledgments
\emph{Acknowledgments---}
We are thankful to Stephen R. Green, Cecilia Fabbri, and Sam Patrick for valuable feedback and fruitful discussions. 
L.\,S. and S.\,W. gratefully acknowledge the support of the Leverhulme Research Leadership Award (RL-2019-020). 
S.\,H.\,V. acknowledges funding from the Deutsche Forschungsgemeinschaft (DFG): Project No. 386119226. 
The authors acknowledge the Institute for Fundamental Physics of the Universe (IFPU) in Trieste for financial support of the focus week program "Analog Gravity Meets Gravitational Waves" which helped in finalizing the project.

\bibliographystyle{apsrev4-2}
\bibliography{biblio}

\appendix

\section{P\"oschl-Teller Green's function}\label{app:green}

In the following, we study the Green's function associated with the wave equation with a P\"oschl-Teller potential with partially-reflective boundary conditions,
\begin{align}
    \partial_x^2 \phi + (\omega^2 -\cosh^{-2}(    \alpha x))\phi = 0\,.
\end{align}
The exact solution is given in term of Hypergeometric functions $_2F_1(a,b,c;y)$ as
\begin{equation}
   \phi =  A\,F(-\omega,y)z^{+i\omega /2\alpha} + 
    B\,F (\omega,y)z^{-i\omega /2\alpha}\,,
\end{equation}
where $y = (1+\exp(-2\alpha x))^{-1} \in (0,1)$ and
\begin{align}
    z = \left(\frac{1-y}{y}\right),\quad l = \frac{1}{2} + \frac{i}{2}\sqrt{\frac{4}{\alpha^2}-1}\,,
\end{align}
and 
\begin{align}
_2F_1(l,\bar{l},1+i\omega;y) = F(\omega,y) \,.
\end{align}
Note that $\lim_{x\to\pm\infty} z(y) = \exp(-2\alpha x)$, which makes the $x\to \infty$ asymptotic solution a superposition of out-going and in-coming waves
\begin{align}
    \phi \sim A  e^{-i\omega x} + B e^{+i\omega x}\,.
\end{align}
By using the reflection transformation of the hypergeometric functions, we can write
\begin{align}
F(\omega,y) =z^{i\omega/\alpha}\, \mathcal{C}_\omega\, F(\omega,1-y) + \mathcal{D}_\omega\, F(-\omega,1-y)\,,
\end{align}
where
\begin{align}
    \mathcal{C}_\omega = \frac{i\omega\Gamma(i\omega)^2}{\Gamma(l+i\omega)\Gamma(\bar{l}+i\omega)}\,,\quad
    \mathcal{D}_\omega = i\omega\frac{|\Gamma(i\omega)|^2}{|\Gamma(l)|^2}\,.
\end{align}
Note that $|\mathcal{C}|^2-|\mathcal{D}|^2 = 1$. 
In this way, we can express the field as
\begin{align}
\phi = z^{+i\omega/2\alpha}\left(A \mathcal{C}_{-\omega} + B \mathcal{D}_\omega \right)F(\omega,1-y) + \nonumber \\
z^{-i\omega/2\alpha}\left(A \mathcal{D}_{-\omega} + B \mathcal{C}_\omega \right)F(-\omega,1-y)\,.
\end{align}

Now, in the limit $x\to \infty$, one has
\begin{align}
    \phi \sim (A \mathcal{C}_{-\omega} + B\mathcal{D}_{\omega}) e^{-i\omega x} +(A \mathcal{D}_{-\omega} + B\mathcal{C}_{\omega}) e^{+i\omega x}\,.
\end{align}
Given the two expressions, we can find the scattering coefficients $R(\omega), T(\omega)$ associated with the potential, from the asymptotic behaviour
\begin{align}
\label{scatt}
    \phi\propto \begin{cases}
    e^{-i\omega x} + R(\omega)\,e^{i\omega x} & ,\,\,x\to +\infty\,, \\
    T(\omega) e^{-i\omega x}&,\,\,x\to -\infty\,,
    \end{cases} 
\end{align}
which gives
\begin{align}
    T = \frac{1}{\mathcal{C}_{-\omega}}\,, \quad R = \frac{\mathcal{D}_{-\omega}}{\mathcal{C}_{-\omega}}\,,
\end{align}
with $|T|^2 + |R|^2 = 1$. 

To evaluate the Green's function, we need to find two solutions, each satisfying the boundary condition in one of the boundaries. In particular, for Eq.~\eqref{BCs}, one finds the two solutions $\phi_L,\,\phi_R$, satisfying the boundary condition at $x_l,x_r$ respectively, to be 
\begin{align}
\label{app_PhiLR}
   \begin{bmatrix}
       \phi_L\\
       \phi_R
   \end{bmatrix}
   \propto z^{+i\omega /2}F(-\omega,y) +
   \begin{bmatrix}
       \bar{\varepsilon}_l\\
       K
   \end{bmatrix}
   z^{-i\omega /2}\,F(\omega,y)\,,
\end{align}
where 
\begin{align}
    \bar{\varepsilon}_l = \varepsilon_l e^{-2i\omega x_l }\,,\quad\bar{\varepsilon}_r = \varepsilon_r e^{2i\omega x_r }\,,
\end{align}
and 
\begin{align}
    K = \frac{T^*}{T}\frac{R\,\bar{\varepsilon}_r-1}{R^*-\bar{\varepsilon}_r}\,.
\end{align}

Furthermore, by choosing the normalisation of Eq.~\eqref{app_PhiLR} such that
\begin{align}
    \phi_L(x\ll -1) = e^{-i\omega x} + \bar{\varepsilon}_l e^{i\omega x} \nonumber \\
    \phi_L(x\gg 1) = \bar{\varepsilon}_re^{-i\omega x} + e^{i\omega x}\,,
\end{align}
the Wronskian $W = \phi_L\phi_R^\prime -\phi_L^\prime \phi_R $ reads
\begin{align}
    W = \frac{2 i \omega}{T}\left(R(\bar{\varepsilon}_r + \bar{\varepsilon}_l) + \frac{T}{T^*}\bar{\varepsilon}_r\bar{\varepsilon}_l-1\right)\,.
\end{align}
Then, the Green's function presented in the main text and plotted in Fig~\ref{fig:spectrum} is given by
\begin{align}
    G(x,x_s, \omega) = \frac{\phi_L(x_s)\phi_R(x)}{W}\,.
\end{align}

\section{Numerical simulations}\label{app:num}

We solve the stochastic wave equation Eq.~\eqref{eq:WaveEq} using the method of lines \cite{Solidoro:2024yxi}. First we define a spatial grid $x_k$ with $k=1...N$ and \mbox{$h = x_{k+1}-x_k$}, spanning from $x_l$ to $x_r$. Here we choose $N = 251$. Then, we construct a 3-point centred difference stencil for the second derivative,
\begin{equation}
    \partial_x^2\psi_k = \frac{\psi_{k+1}-2\psi_k+\psi_{k-1}}{h^2}\,.
\end{equation}
Introducing $\Pi = \partial_t\psi$,  Eq.~\eqref{eq:WaveEq} can be solved as
\begin{equation} \label{num_eq}
    \partial_t \begin{pmatrix}
        \psi \\ \Pi
    \end{pmatrix} = \begin{pmatrix}
        0& \mathds{1}_\mathrm{N} \\ {}^\mathrm{N}\partial_x^2-V & B
    \end{pmatrix}\begin{pmatrix}
        \psi \\ \Pi
    \end{pmatrix}
         +     \begin{pmatrix}
        0 \\ I
    \end{pmatrix}\,,
\end{equation}
where $\mathds{1}_N$ is the $ N \times N$ identity matrix, $ {}^\mathrm{N}\partial_x^2$ is the second-order derivative expressed on a $N$-point mesh and  $B = -(2/h)\mathrm{diag}_N(\mu_l,0,...0,\mu_r)$ is a diagonal matrix with non-zero entries in the top-left and bottom-right corner implementing the boundary conditions. Specifically, Eq.~\eqref{BCs} are obtained choosing
\begin{align}
    \mu_{r/l} = \frac{1-\varepsilon_{r/l}}{1+\varepsilon_{r/l}}\,.
\end{align}

For the Shallow-water equation Eq.~\eqref{SWeq}, the evolution matrix becomes
\begin{equation} \label{num_eq_SW}
    \partial_t \begin{pmatrix}
        \psi \\ \Pi
    \end{pmatrix} = \begin{pmatrix}
        -i mC/r^2 & \mathds{1}_\mathrm{N} \\ {}^\mathrm{N}\partial_x^2-U & B-i mC/r^2
    \end{pmatrix}
    \begin{pmatrix}
        \psi \\ \Pi
    \end{pmatrix}
     +     \begin{pmatrix}
        0 \\ I
    \end{pmatrix}\,.
\end{equation}

The field is evolved in time using a 4-th order Runge-Kutta algorithm, starting from an empty array as initial data. To mimic an incoherent source, each realisation is driven by a Gaussian white-noise signal injected at a fixed point $x_s$ located at 5 grid-points from the left boundary. At each time step, the source is re-sampled from a Gaussian distribution with variance $\sigma$. We then extract the time series at a fixed grid-point $x_d$ at 20 grid-points right boundary. After discarding the initial transient, we compute the Fourier transform of the late-time signal. 

The training set is generated by drawing the parameters $\theta$'s from uniform distributions and extracting the spectrum. The resulting spectra are then used as inputs to train the neural network. In particular, we use the Neural Posterior Estimator (NPE) of the \texttt{sbi} package~\cite{tejero-cantero2020sbi,tejero_cantero_2022_8192532,BoeltsDeistler_sbi_2025} with a masked autoregressive flow (MAF) as density estimator, a normalizing-flow architecture that represents the posterior as a sequence of invertible, autoregressive transformations of a multivariate Gaussian base distribution \cite{papamakarios2017masked}. For the cases presented here, the training set consists of $10^5$ spectra for the P\"oschl-Teller case and $10^6$ for the shallow-water model. In both cases, further $10^3$ independent realisations are generated as the testing set. 
Once trained, the resulting posterior can be sampled to obtain parameter constraints for a given spectrum.

\begin{figure*}
    \centering
    \includegraphics[width=1.0\linewidth]{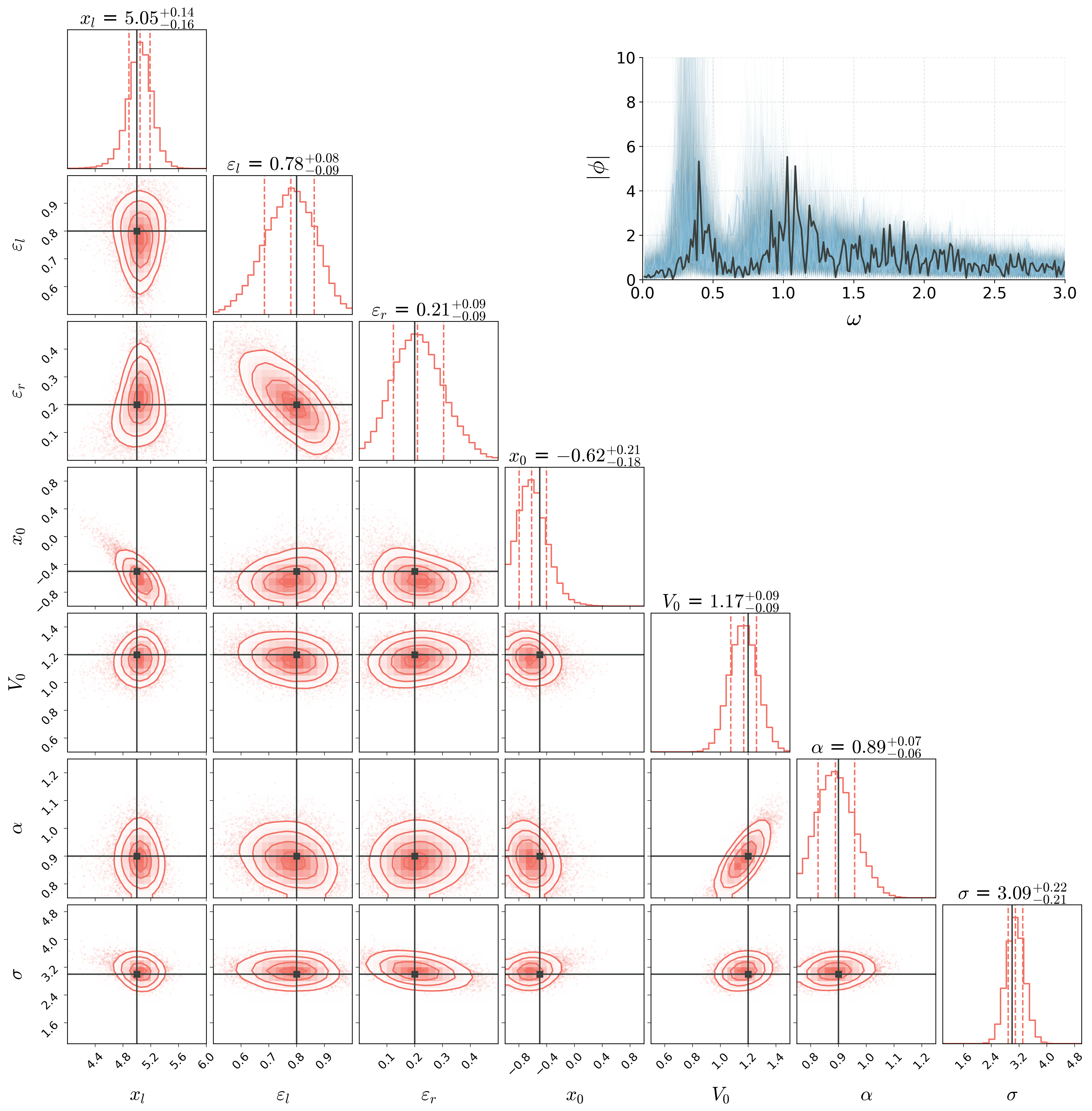}
    \caption{Corner plot showing a representative example of applying NPE to a simulated observation of a Pöschl–Teller analogue black hole. 
    Shown is the posterior distribution of conditioned on a single simulated spectrum with injected parameters $\theta_\mathrm{PT} = [1.2, 0.9, -0.5, -5, 0.8, 0.2, 3]$ (black lines).
Diagonal panels display the one-dimensional marginal posteriors with median and 68\% credible intervals, while off-diagonal panels show the corresponding joint distributions.
Inset: observed spectrum used for inference (black) compared to a subset of spectra from the training set (blue).}
\label{fig:PT_corner}
\end{figure*}

\begin{figure}
    \centering
    \includegraphics[width=1\linewidth]{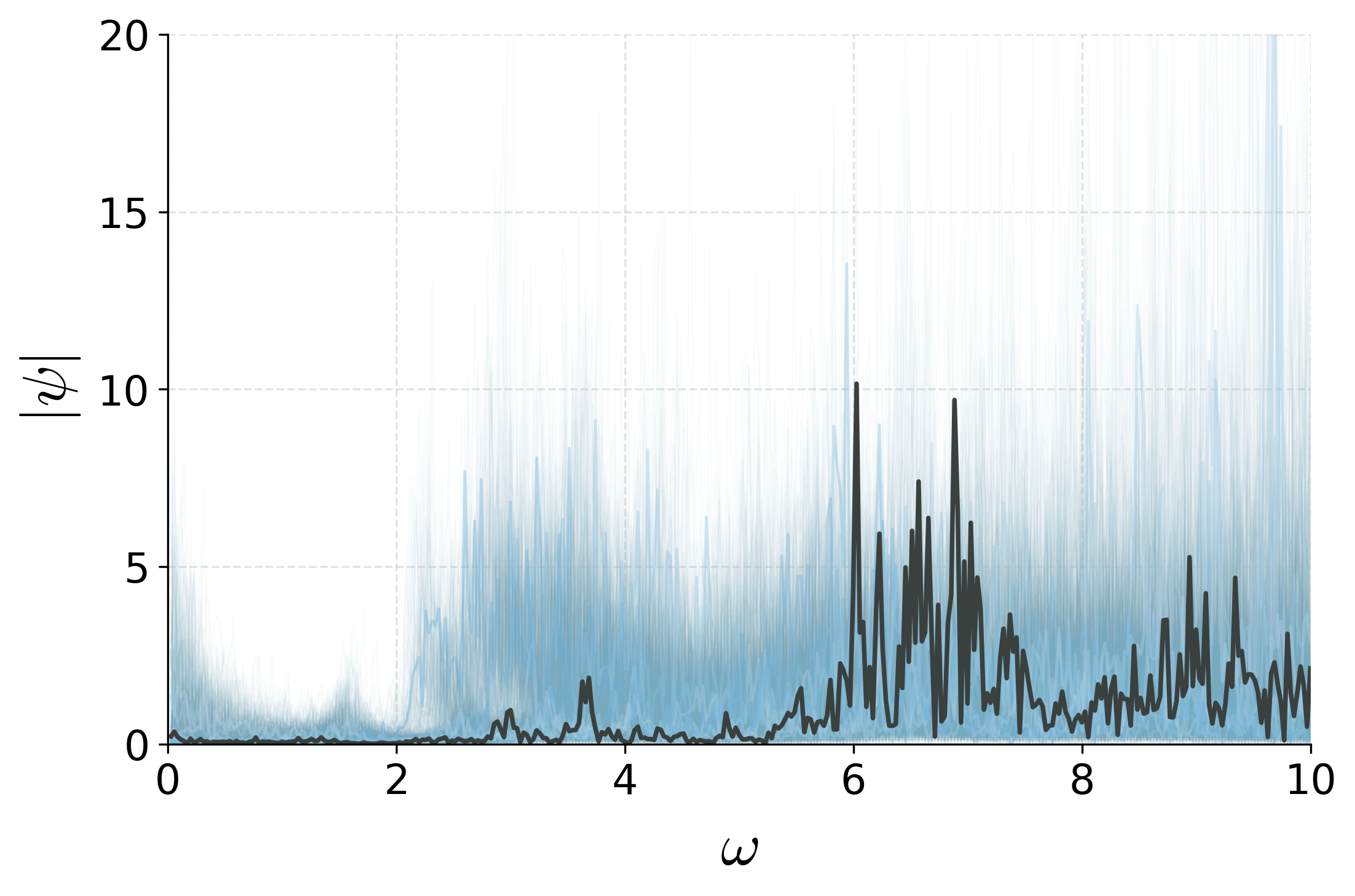}
    \caption{Spectrum used in the application shown in Fig.~\ref{fig:SW_Corner} (dark curve) plotted over a subset of spectra used in the NPE training set (blue curves). }
    \label{fig:SW_Train}
\end{figure}

In Fig~\ref{fig:PT_corner}, we show the result of the NPE applied to the noisy spectrum presented in Fig~\ref{fig:spectrum}. 
The axis shows the full prior ranges of all parameters. 
We find that the reflectivities $\varepsilon_{r/l}$ are the least constrained parameters (compared to their priors), and the joint distribution shows strong correlation between the two. 
The inset shows the injected spectrum (black line) together with a sample from the training set (blue lines), showcasing its large variability. 
The same comparison in the shallow-water case is shown in Fig~\ref{fig:SW_Train}, where the dark curve represents the same spectrum as used in Fig~\ref{fig:SW_Corner}.

\begin{figure*}
    \centering
    \includegraphics[width=1.0\linewidth]{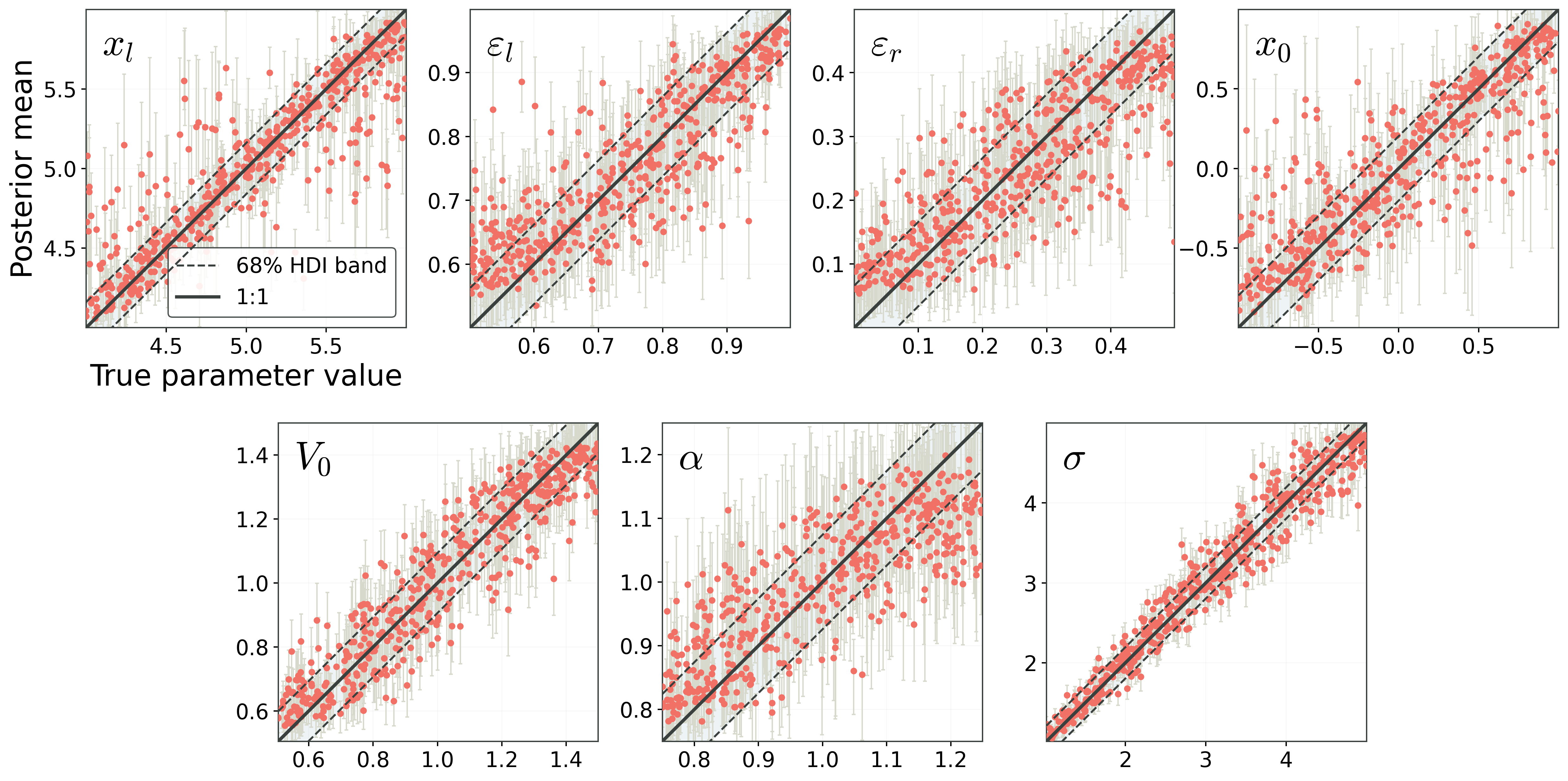}
    \caption{Validation of the NPE for P\"oschl-Teller perturbations using 500 samples from the test set. 
    The mean of the posterior is plotted against its corresponding true value (orange markers), with 68\% highest-density interval (HDI) error bars shown for every sample. 
    The solid diagonal indicates perfect recovery, while the dashed lines denote the average 68\% HDI width around the identity.}
    \label{fig:PT_Valid}
\end{figure*}

In Fig~\ref{fig:PT_Valid}, we compare the prediction from the NPE to the injected values of the test set for 500 different P\"oschl-Teller spectra. 
\section{Simulation-based calibration}\label{app:calibration}
\begin{figure*}
    \centering
    \includegraphics[width=0.49\linewidth]{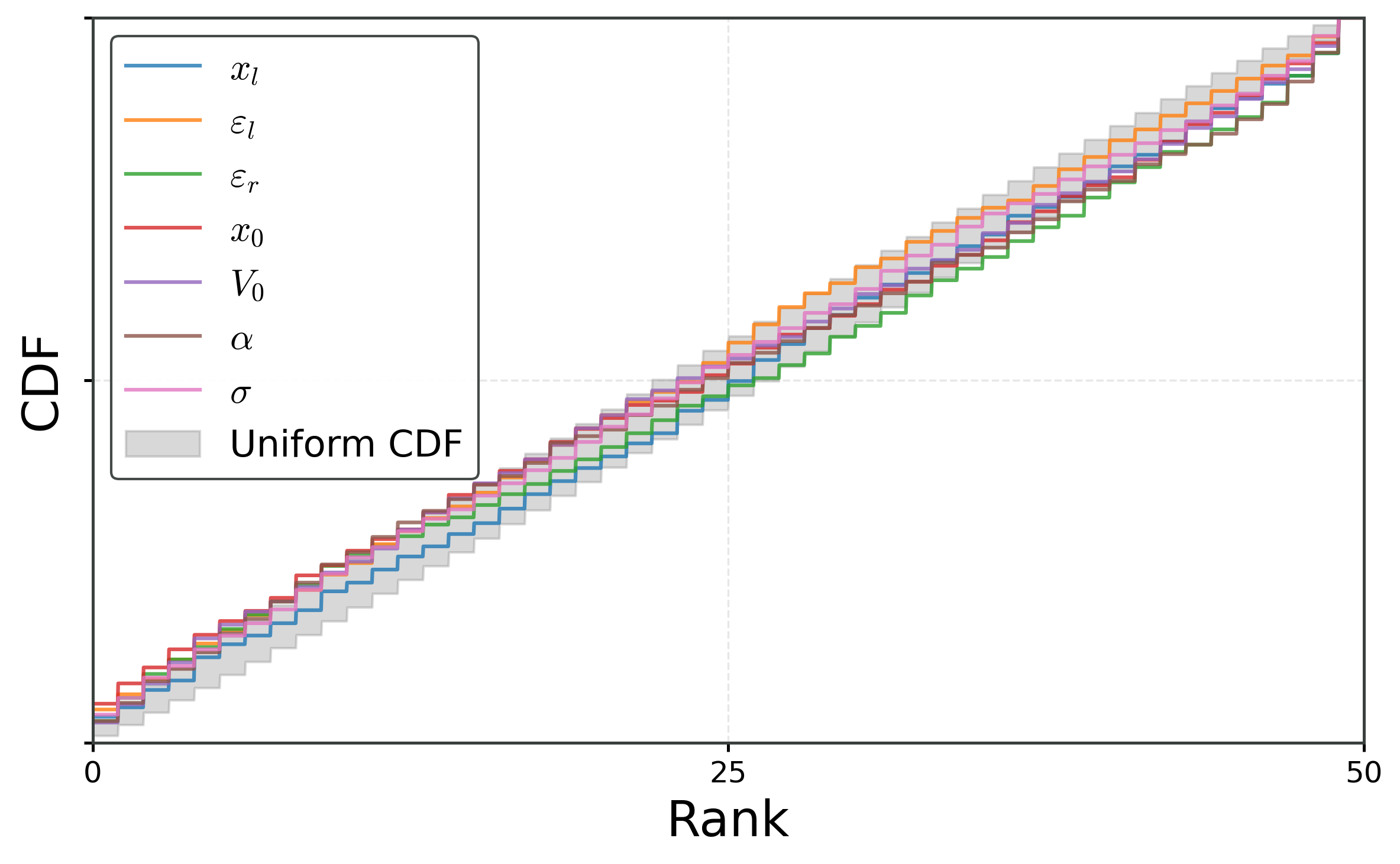}
    \hfill \includegraphics[width=0.49\linewidth]{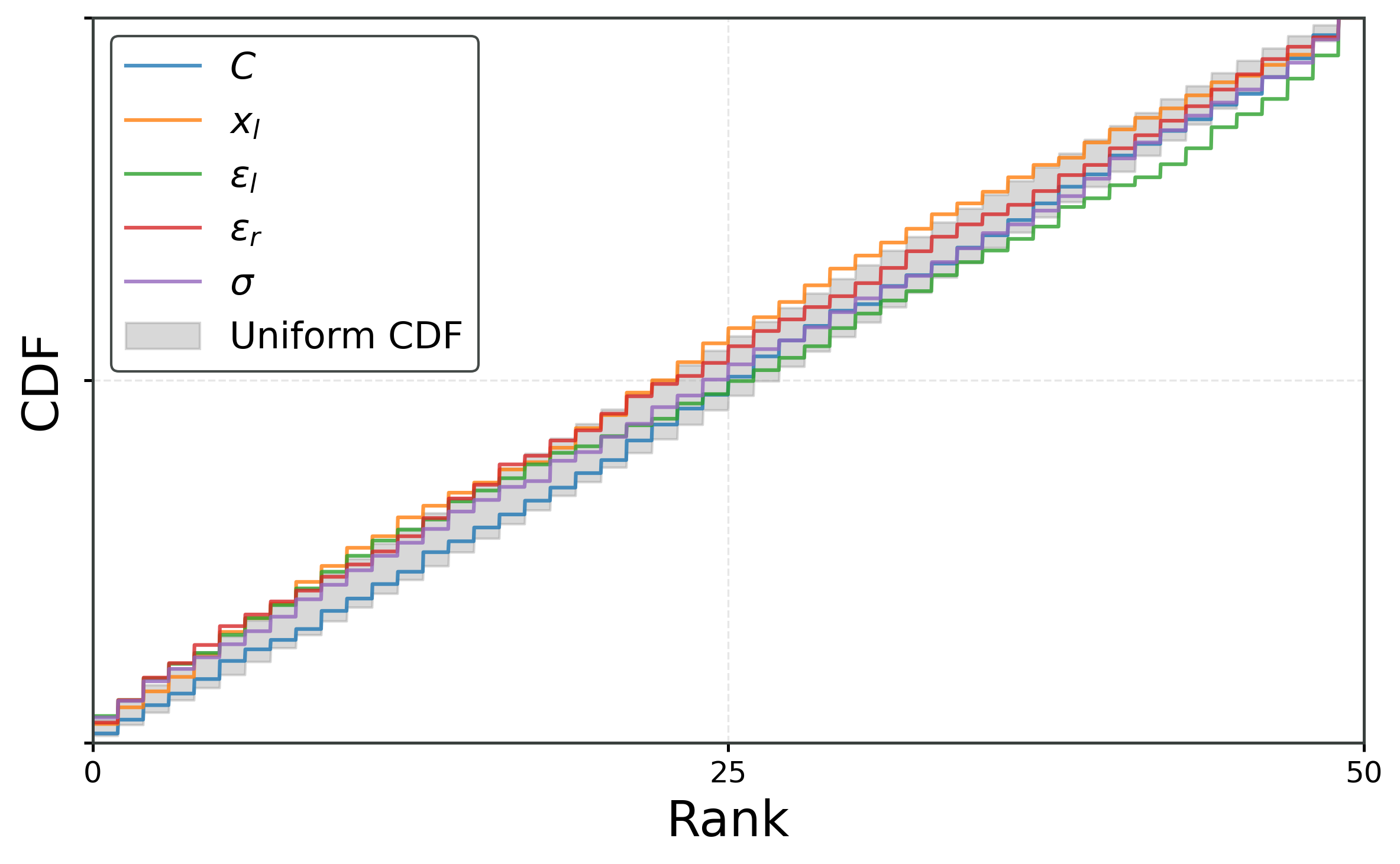}
    \caption{Simulation-based calibration (SBC) using rank statistics for the P\"oschl-Teller case (left panel) and the shallow-water case (right panel). 
    }
    \label{fig:Ranks_cdf}
\end{figure*}

We assess the calibration of the NPE using standard simulation-based calibration (SBC) diagnostics, as implemented in the \texttt{sbi} package~\cite{tejero-cantero2020sbi,tejero_cantero_2022_8192532,BoeltsDeistler_sbi_2025}. In particular, we employ rank statistics and their cumulative distribution functions (CDFs). For a well-calibrated posterior, the resulting rank statistics are expected to be uniformly distributed. Deviations from uniformity in the rank histograms or their empirical CDFs indicate biased inference or misestimated posterior uncertainties.
In Fig.~\ref{fig:Ranks_cdf}, we show the ranks of each parameter versus its CDF of the P\"oschl-Teller and shallow-water case, respectively. 
The rank statistics suggest that the NPE is well calibrated, with a better performance for the shallow-water case, which one may expect since the latter one has fewer parameters. 
Together with the results reported in Fig.~\ref{fig:SW_Valid}, we conclude that NPE is a very valuable tool for analysing noisy analogue systems. 

\end{document}